\documentclass{optica-article}

\journal{opticajournal} 

\articletype{Research Article}
\usepackage{lineno}

\begin{document}

\title{Laser Pulse Diagnostics of Ultrafast <8 fs Pulses Through Two-Photon Absorption Fluorescence In Liquid Media - The Role of GVD and Third-Order Dispersion}

\author{Asad Mahmood\authormark{*}\authormark{1}, Ra\'ul Puente\authormark{1}, Herman Batelaan\authormark{1}, Rafeeq Syed\authormark{1,2}, and Cornelis J. G. J. Uiterwaal\authormark{1}}

\address{\authormark{1}Department of Physics and Astronomy, University of Nebraska Lincoln, 855 N 16th Street, Lincoln, NE 68588,
USA}
\address{\authormark{2}Now at ASML, Wilton, Connecticut}

\email{\authormark{*}amahmood9@huskers.unl.edu} 

\begin{abstract*} 
This study investigates the propagation of an ultrafast laser pulse through a liquid medium. A femtosecond laser oscillator with a pulse duration of less than 8 fs is used. By conducting experiments with coumarin and fluorescein dyes in water, methanol, and chloroform, we analyze two-photon absorption (TPA) fluorescence, a method pioneered by Schr\"{o}der [Opt. Express \textbf{14}, 10125 (2006)]. A numerical algorithm we developed to model the fluorescence signal determines the group velocity dispersion (GVD), the third-order dispersion (TOD), and the group delay dispersion (GDD). Autocorrelation measurements combined with a detailed analysis confirm the validity of our method and the accuracy of the retrieved temporal profile of the pulse. This cost-effective approach is robust and useful for laser pulse characterization, outperforming traditional methods in terms of alignment sensitivity. Our method allows us to study the time evolution of the pulses as they propagate through the liquid, determines higher-order phase terms as acquired by the pulse while reflecting off the chirped mirrors and propagating through the liquid, and even works for non-Gaussian spectral intensities of the laser.
\end{abstract*}

\section{Introduction}
Ultrafast optics is a rapidly advancing field that relies on the generation, manipulation, and characterization of ultrashort laser pulses, typically ranging from femtoseconds (10\(^{-15}\) seconds) to picoseconds (10\(^{-12}\) seconds)\cite{trebino2020highly}. This field has revolutionized scientific and technological domains by enabling the study and control of phenomena occurring on timescales comparable to atomic and electronic motions within molecules.

Accurate characterization of ultrashort laser pulses is crucial for both scientific research and technological applications. Ultrashort pulses facilitate the study of rapid processes, such as chemical reactions, electron dynamics, and light-matter interactions on femtosecond timescales\cite{trebino2020highly,guo2020ultrashort}. Accurate pulse characterization is essential for interpreting experimental results and gaining insights into these ultrafast phenomena. Many applications, including precision material processing, ultrafast spectroscopy, and attosecond science, rely on generating and controlling ultrashort pulses with specific temporal and spectral properties\cite{trebino2020highly,franco2023curve}. Experimental data from ultrashort pulse characterization techniques provide valuable information for validating and refining theoretical models describing pulse generation, propagation, and interaction with matter\cite{trebino2020highly}. By accurately characterizing ultrashort laser pulses, researchers and engineers can unlock new frontiers in scientific exploration, technological innovation, and the development of advanced applications across various domains\cite{trebino2020highly,Trebino2000,753647}.

Numerous techniques have been developed for characterizing ultrashort laser pulses. Among these, the commercially available methods capable of resolving few-cycle sub-femtosecond laser pulses include Row Optical Autocorrelator (ROC)\cite{ROCAutoc14:online}, Frequency-Resolved Optical Gating (FROG)\cite{676498, Singlesh34:online}, Spectral Phase Interferometry for Direct Electric-field Reconstruction (SPIDER)\cite{witting2011characterization}, Multiphoton Intrapulse Interference Phase Scan (MIIPS)\cite{dantus2007miips}, and Dispersion Scan (D-scan)\cite{fabris2015single, miranda2012characterization}. These techniques rely on nonlinear interactions, which make them sensitive to alignment, require high laser power, and necessitate the use of wavelength-dependent nonlinear crystals. These prominent methods are particularly well-suited for characterizing ultrashort, few-cycle pulses, including those as short as sub-femtosecond.

In this paper we discuss our pulse characterization method, which is less sensitive to alignment and collimation, more cost-effective compared to commercial devices, and highly effective for sub-femtosecond pulses (below 8 fs) across a broad wavelength range exceeding 200 nm. This new method builds on the previous successful experiments performed by Théberge et al.\cite{Theberge:06}, Scarborough et al.\cite{scarborough2008measurements}, and Syed et al.\cite{syed2023ultrafast} (see also ~\cite{syed2023tale}). However, unlike this previous work, we use a numerical analysis instead of an analytical approach. This means that we no longer have to assume a perfectly Gaussian shaped pulse. In addition, our method provides a way to minimize the pulse duration, to determine the intrinsic group delay dispersion (GDD) of the laser, and to study higher-order phases of the liquid media.

This paper is organized as follows. First, we present the theory of ultrafast pulse propagation, including the numerical algorithm. Next, our experiment is discussed. Briefly, a negatively chirped laser pulse leaves the source, enters a liquid medium with dissolved fluorescent dye, resulting in a two-photon absorption (TPA) fluorescence signal which is recorded photographically. Finally, we discuss our findings. The group velocity dispersion (GVD) of the three liquids is verified, the GDD of the laser and the third-order dispersion (TOD) of the liquid are found. The laser's pulse duration that depends on the GVD, GDD and TOD is found and confirmed to match an independent non-linear autocorrelation measurement and the manufacturer's specifications of the laser. 

\section{Theory}

For a given spectral intensity of pulsed lasers, denoted as \( I(\omega) \), the complex amplitude of the electric field $\tilde{U}(z, \omega)$ at a specific position \( z \) within a dispersive medium can be expressed as\cite{Trebino2000,syed2023tale,marton2022study,diels2006ultrashort,siegman1986lasers,agrawal2007nonlinear}:

\begin{equation}
\tilde{U}(z,\omega) = \tilde{U}(0,\omega) e^{i(\phi_\text{pulse}(\omega) + \phi_\text{medium}(\omega,z))}
\label{1.0}
\end{equation}

Here, \( \tilde{U}(0,\omega) = \sqrt{I(0,\omega)} \) represents the complex amplitude at the initial position, while \( \phi_\text{pulse}(\omega) \) and \( \phi_\text{medium}(\omega,z) \) account for the phase contributions from the pulse and the medium, respectively. In particular, \( \phi_\text{pulse}(\omega) \) captures phase distortions introduced by pulse compressors such as gratings, prism pairs, chirped mirrors (CM), or any inherent phase generated as a result of amplification within the laser system, among other factors that influence the overall pulse shape. On the other hand, $\phi_\text{medium}(\omega,z) = \beta(\omega)z$ represents the phase accumulated due to dispersion by the medium over the propagation distance $z$. $\beta(\omega)$ and $\phi_\text{pulse}(\omega)$ are Taylor expanded around the carrier frequency $\omega_0$ as\cite{Trebino2000,syed2023tale,marton2022study,diels2006ultrashort,siegman1986lasers,agrawal2007nonlinear}:
\begin{equation}
\beta(\omega) = \sum_{n=0}^\infty \frac{\beta^{(n)}(\omega_0)}{n!}(\omega-\omega_0)^n
\label{1.1}
\end{equation}
\begin{equation}
\phi_\text{pulse}(\omega) = \sum_{n=0}^\infty \frac{\phi_n}{n!}(\omega-\omega_0)^n
\label{1.2}
\end{equation}

The coefficients $\beta^{(1)}$, $\beta^{(2)}$, etc., correspond to the inverse group velocity, GVD, TOD and higher order dispersion, respectively. Similarly, $\phi_1$, $\phi_2$, etc., describe the group delay (GD), GDD, and higher-order terms associated with the pulse phase. The pulse envelope profile in the time domain can be obtained by taking the inverse Fourier transform of Eq.~(\ref{1.0}):
\begin{equation}
U(z,t) = \mathcal{F}^{-1}\{\tilde{U}(z,\omega)\}
\label{1.3}
\end{equation}
where $\mathcal{F}^{-1}$ denotes the inverse Fourier transform operation.

\subsection{Pulse Compression and Pulse Duration Broadening}
Consider a scenario where $\phi_\text{pulse}(\omega)$ is manipulated such that as the pulse propagates into the medium, there exists a point $z=z_0$ where $\phi_\text{medium}(\omega, z_0)$ approximately cancels $\phi_\text{pulse}(\omega)$, yielding $U(z_0,t) = U(0,t)=\mathcal{F}^{-1}\{\tilde{U}(0,\omega)\}$, which is known as the transform-limited pulse. As an illustrative example, let us consider a perfect Gaussian pulse envelope, and for mathematical simplicity, we limit $\phi_\text{medium}(\omega, z)$ to $\beta_2(\omega)\cdot z$ and $\phi_\text{pulse}(\omega)$ to $\phi_2$, corresponding to the GVD and GDD terms, respectively. Eq.~(\ref{1.0}) becomes 
\begin{equation}
\tilde{U}(z,\omega) = e^{-2\ln2\left(\frac{\omega-\omega_0}{\Delta\Omega}\right)^2} e^{i\frac{(\omega-\omega_0)^2}{2}(\text{GDD} + \text{GVD}\cdot z)}
\label{2}
\end{equation}
where $\Delta\Omega$ is the full-width-half-maxima (FWHM) spectral bandwidth of the laser.
Note that pulse compressors typically introduce a negative GDD value, while most dispersive media exhibit positive GVD values. Equation~(\ref{2}) is used to extract the pulse duration $\Delta T(z)$ of the spectral electric field as a function of $z$ \cite{syed2023ultrafast,scarborough2008measurements,Theberge:06,syed2023tale}:
\begin{equation}
\begin{split}
\Delta T(z) = \Delta T_\text{opt} \sqrt{1 + \left(\frac{4 \ln 2 \cdot (\text{GVD}\cdot z+\text{GDD})}{\Delta T_\text{opt}^2}\right)^2}\\ = \Delta T_\text{opt} \sqrt{1 + \left(\frac{4 \ln 2 \cdot \text{GVD}\cdot (z-z_{0})}{\Delta T_\text{opt}^2}\right)^2}
\label{3}
\end{split}
\end{equation}
where $\Delta T_\text{opt}$, being the FWHM, is the transform-limited pulse duration given by $\Delta T_\text{opt}\cdot\Delta\Omega\approx2.77$ (for a Gaussian pulse), which occurs at $z=z_{0}=-\text{GDD}/\text{GVD}$ within the medium.
\subsection{TPA Fluorescence Intensity}
TPA fluorescence is a nonlinear optical process that involves the simultaneous absorption of two photons by a fluorescent dye or fluorophore\cite{rumi2010two}. By carefully selecting fluorescent dyes with large enough TPA cross-sections and having a TPA absorption spectrum that matches the spectrum of the laser pulse in use, one can leverage the TPA fluorescence signal to study the properties of the laser pulse itself, such as its pulse duration. The TPA fluorescence intensity $S(z)$ is proportional to the fourth power of the temporal amplitude $U(z,t)$ profile of the laser pulse given as\cite{syed2023ultrafast,scarborough2008measurements,Theberge:06,syed2023tale}:
\begin{equation}
S(z) \propto \int_{-\infty}^{\infty} |U(z,t)|^4 \, \mathrm{d}t
\label{4}
\end{equation}

When a negatively chirped Gaussian pulse, as given by Eq.~(\ref{2}), with a known GDD value enters a fluorescence enabled medium with a known GVD, one can estimate the pulse-duration from the TPA fluorescence intensity $S(z)$ using the following expressions\cite{syed2023ultrafast,scarborough2008measurements,Theberge:06,syed2023tale}:
\begin{equation}
S(z) \propto \frac{\text{exp}(-2\alpha z)}{\Delta T_\text{opt} \sqrt{1 + \left(\frac{4 \ln 2 \cdot \text{GVD}\cdot (z-z_{0})}{\Delta T_\text{opt}^2}\right)^2}}
\label{5}
\end{equation}
and
\begin{equation}
\Delta T_\text{opt} = \sqrt{\frac{2\ln2}{\sqrt{3}}z_{1/2}\times \text{GVD}}
\label{6}
\end{equation}
where $\alpha$ is the linear absorption coefficient of the medium, and $z_{1/2}$ denotes the longitudinal spatial FWHM of the TPA fluorescence signal $S(z)$.
\begin{figure}[ht!]
    \centering
    \includegraphics[width=0.7\columnwidth]{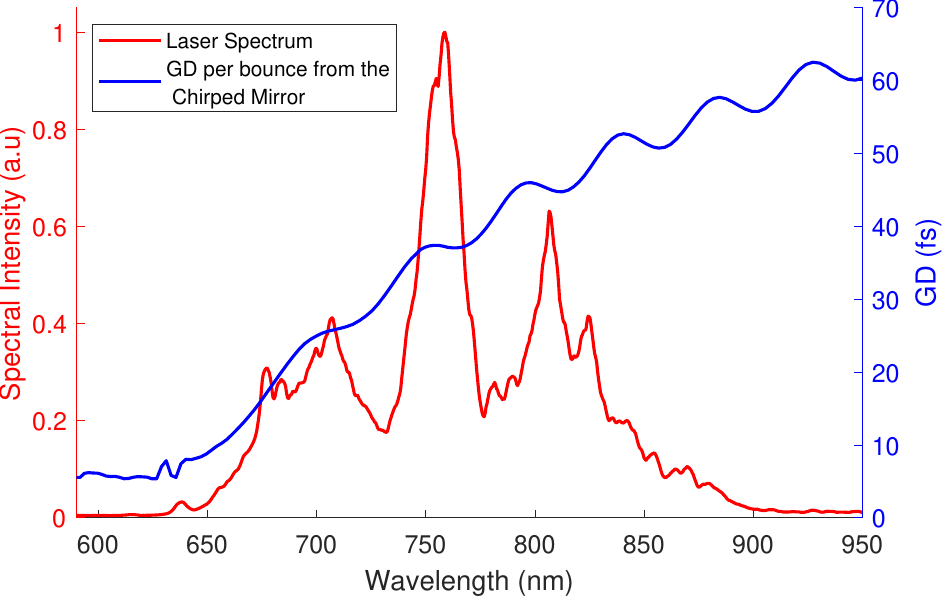}
    \caption{\textbf{Red}: Non-Gaussian spectral intensity $|\tilde{U}(0,\omega)|^2$ of Thorlabs' Octavius femtosecond laser\cite{TiSapphi70:online} measured by the spectrometer. \textbf{Blue}: Wavelength dependence of GD$_\text{per bounce}$ data of the CMs obtained from Thorlabs\cite{ChirpedM64:online}, showing an overall nonlinear curve in addition to oscillatory behavior.}

    \label{spectra_GD_combined}
\end{figure}
\subsection{Limitations for Broadband, Non-Gaussian Pulses}
For wide-band ultrashort lasers, the inclusion of higher-order dispersion terms beyond GVD becomes crucially important \cite{li2015fourth} because GVD is no longer constant over the entire spectral range of broadband lasers. Furthermore, pulse compressors like CMs, due to the way they are manufactured, do not maintain a constant GDD (or linear GD) across the spectral range of such short pulses, as illustrated in Fig.~\ref{spectra_GD_combined}. Consequently, Eq.~(\ref{3}), Eq.~(\ref{5}), and Eq.~(\ref{6}) are no longer valid for retrieving the exact pulse duration information using the TPA fluorescence method. Nevertheless, in this paper, we demonstrate a novel method of determining the pulse duration by efficiently including the higher-order effects to simulate results very close to experimental observations.

\subsection{Simulation}
\label{simulation}
We developed a MATLAB script to simulate the experiment by incorporating the measured spectral intensity $|\tilde{U}(0,\omega)|^2$ and the group delay per bounce ($\text{GD}_{\text{per bounce}}$) provided by the CM manufacturer, as shown in Fig.~\ref{spectra_GD_combined}. The spectral amplitude \( \tilde{U}(z,\omega) \) is calculated as the pulse propagates a distance \( z \) within the medium. We assume the pulse entering the medium is not transform-limited and contains at least a second-order background phase. Additionally, the GVD and TOD of the medium are considered to find the propagating spectral amplitude (see Eq.~(\ref{1.0}), Eq.~(\ref{1.1}), and Eq.~(\ref{1.2}))

\begin{equation}
\begin{split}
\tilde{U}(z,\omega)=\tilde{U}(0,\omega)\cdot e^{i\frac{\omega^2}{2}(\text{GDD}_\text{background}+\text{GVD} \times z)}\cdot\\
e^{i\frac{\omega^3}{6}(\text{TOD} \times z)}\cdot e^{i(\text{Bounces}\times\text{GD}_{\text{per bounce}} \times \omega)}
\end{split}
\label{1.01}
\end{equation}

\begin{figure}[ht!]
    \centering
    \includegraphics[width=0.5\columnwidth]{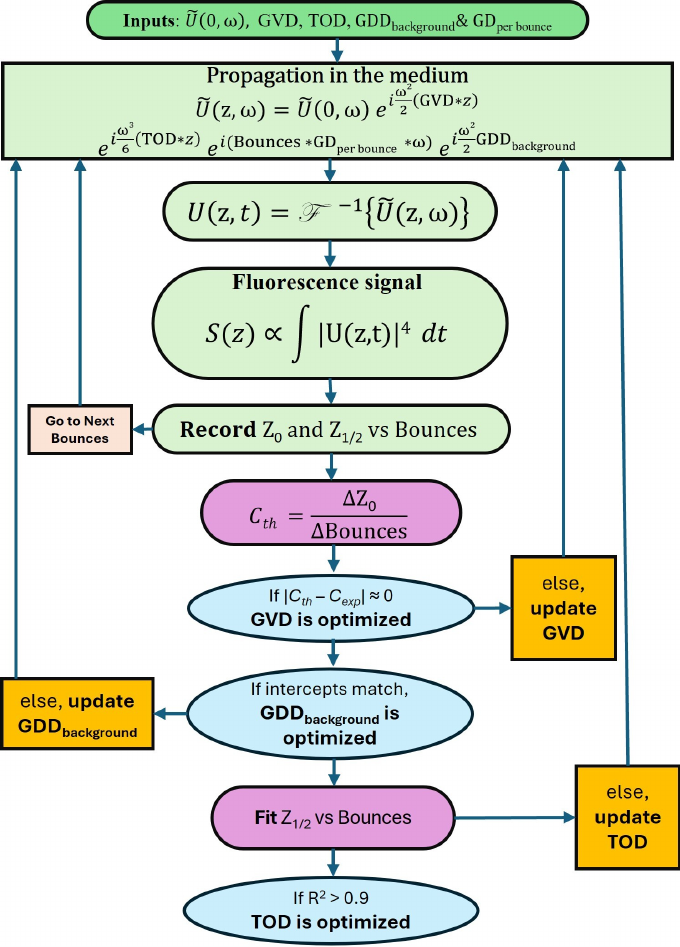}
    \caption{Flowchart for determining the medium parameters (GVD,TOD) and background GDD: The pulse is prepared using the measured spectral intensity data, GD$_{\text{per bounce}}$ of CMs, and trial GVD, TOD and GDD$_\text{background}$ values. The peak fluorescence intensity position $z_0$ is then determined by varying the number of bounces, and the slope $C_\text{th}$ is recorded. This loop is repeated iteratively, updating the GVD value until $C_\text{th} \approx C_\text{exp}$, thus yielding the optimized GVD for that particular liquid medium. In the same iterative fashion, TOD and GDD$_\text{background}$ are optimized.}
    \label{algorithm}
\end{figure}
Following this, \( U(z,t) \) is computed from Eq.~(\ref{1.3}), from which the two-photon absorption (TPA) fluorescence intensity $S(z)$ is obtained (using Eq.~(\ref{4})) and recorded. The algorithm, the flowchart of which is shown in Fig.~\ref{algorithm}, optimizes the following parameters based on the simulated data:

\begin{enumerate}
    \item GVD of the medium by matching the slope of the simulated peak fluorescence intensity position $z_0$ versus Bounces data to the experimentally obtained curves.
    \item GDD$_{\text{background}}$ by matching the $y$-intercept of the simulated peak fluorescence intensity position $z_0$ versus Bounces to the experimentally obtained curves.
    \item TOD of the medium by matching the FWHM values of the simulated $S(z)$ vs $z$ signal to the experimentally obtained curves.
\end{enumerate}
One might argue that this paper primarily focuses on determining the pulse duration, not on the properties of the medium, such as GVD and TOD. However, retrieving accurate pulse information is only possible when the simulated TPA fluorescence signal closely matches the experimental results. Therefore, simulating all three parameters—GVD, GDD$_\text{background}$, and TOD—to match the experimental data was crucial for ensuring that the pulse parameters, especially the pulse duration, are extracted as accurately as possible.

\begin{figure}[ht!]
\centering
\includegraphics[width=0.5\columnwidth]{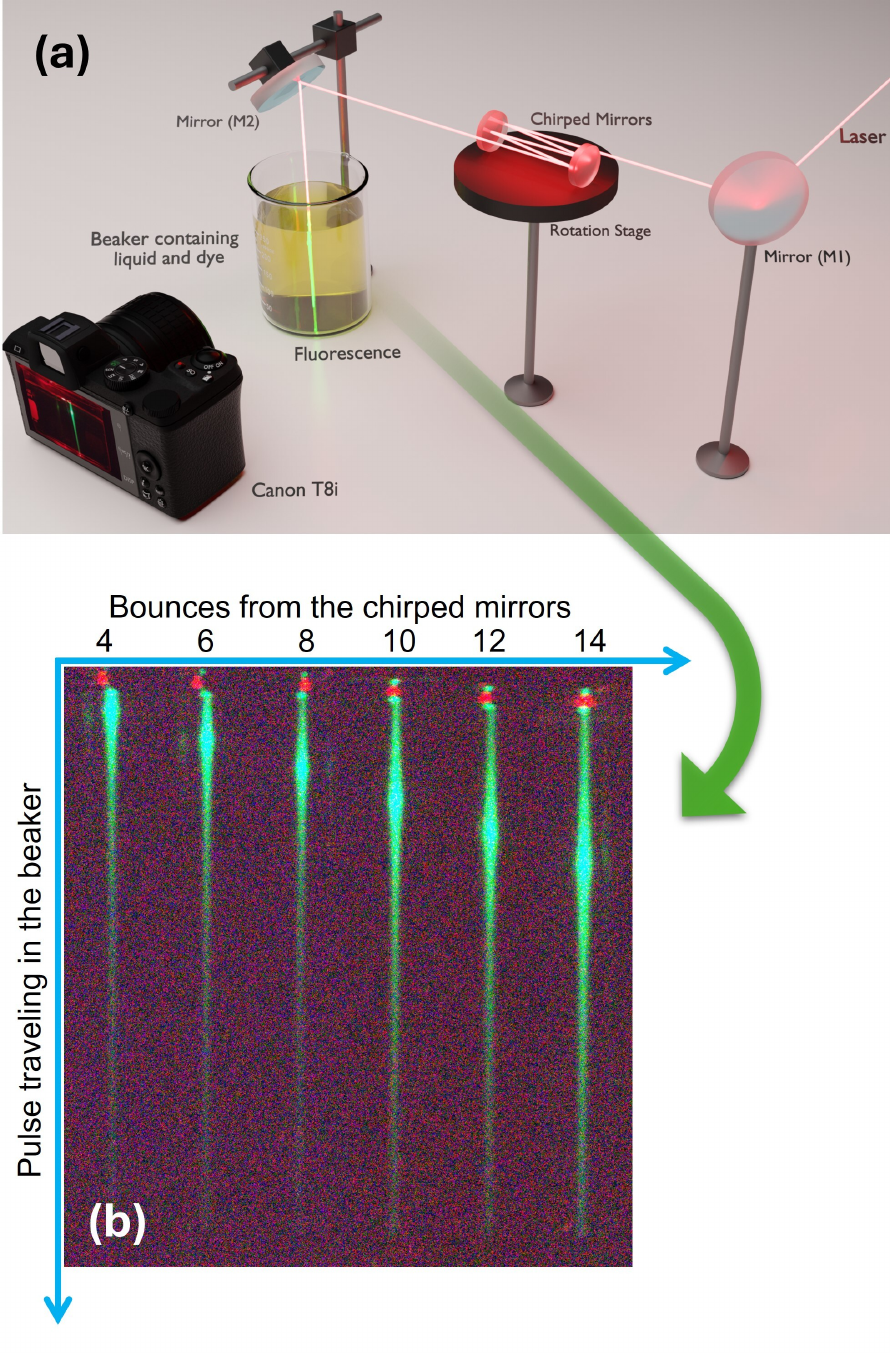}
 \caption{\textbf{(a)}: Experimental setup. The laser beam from the source hits the first broadband mirror M1 for alignment, bounces off a pair of CMs to become negatively chirped, and then reflects off another broadband mirror M2, which ultimately steers the beam straight into a beaker containing the liquid medium, producing a TPA fluorescence signal that is captured by the camera. The amount of chirp depends on the number of bounces, which can be controlled by the rotation stage. \textbf{(b)}: Captured images of the measured TPA fluorescence for the chloroform-coumarin mixture as an example. Note how the peak fluorescence intensity position $z_0$ varies as the number of bounces off the CMs increases.}
    \label{experimental_steup}
\end{figure}
\section{Experimental Procedure}
\subsection{Laser System and the Optical Path}

The experimental setup utilizes a Thorlabs' Octavius femtosecond laser\cite{TiSapphi70:online}, a Ti:sapphire oscillator emitting a broad spectrum of >200 nm, corresponding to a shortest possible pulse duration of <8 fs, as per the manufacturer's specifications. Our primary objective is to experimentally verify this claimed pulse duration.

After exiting the Octavius, the laser beam encounters a broadband mirror (M1) before undergoing multiple reflections from a pair of CMs. All mirrors, including M1, M2, and CMs, were sourced from Thorlabs and covered the full spectral range of the laser. The number of reflections is adjustable by rotating the stage on which the CM pair is mounted. We varied the number of reflections from 4 to 12 in steps of 2, introducing different degrees of negative chirp to the pulse. The setup is visualized in Fig.~\ref{experimental_steup} (a). Following the CM pair, the negatively chirped laser beam is directed onto a second broadband mirror (M2) and then into a beaker containing a liquid medium in which a fluorescent dye is dissolved. Alignment of the beam entering the beaker was ensured using a laser leveler. Special attention was paid to maintaining the beam position away from the edges of the liquid meniscus to prevent distortion. Moreover, the propagation length in the liquid medium (beaker) was significantly shorter than the laser beam's diffraction length.

\subsection{Sample Preparation}
Three different liquid media were utilized:
\begin{itemize}
    \item Coumarin (540A) dye dissolved in chloroform 
    \item Fluorescein dye dissolved in water 
    \item Fluorescein dye dissolved in methanol
\end{itemize}
To minimize dispersion due to the dye, which would compromise the measurement of the dispersive properties of the liquids, the dye concentrations were kept low (less than 0.002 M), as done in similar previous experiments\cite{syed2023ultrafast,scarborough2008measurements,Theberge:06,syed2023tale}. Laser power was maintained below 100 mW to prevent saturation of the fluorescent dye and other nonlinear effects (such as self-focusing, self phase modulation, etc.) during pulse propagation\cite{svelto1974self}. The water was de-ionized and purified. The chloroform and methanol with purity levels exceeding $99.9\%$ were procured from Sigma-Aldrich and used as purchased.

\subsection{Data Acquisition and Processing}

To capture the two-photon absorption (TPA) fluorescence signal, a Canon T8i digital camera was positioned at a sufficient distance from the beaker to avoid parallax distortion. The camera was focused on the fluorescence signal, aligned parallel to a scale, which was later used to convert pixels to length units using ImageJ software \cite{schneider2012nih}. The setup was enclosed in darkness to minimize background light interference. The camera was operated remotely using a MATLAB script, with the exposure time optimized to prevent pixel saturation, especially at peak fluorescence intensity. Raw data in CR3 format were chosen to preserve the integrity of the captured information.
Raw CR3 data were then extracted using the open-source tool dcraw\cite{Decoding63:online} and converted to lossless TIFF format using MATLAB. Since the fluorescence signal from both coumarin and fluorescein dyes emits in the green region of the visible spectrum, only the green channel of the RGB data was selected for further analysis. The fluorescence signal was also normalized by the input power of the beam and the exposure time to ensure comparability across all three liquid media. All other data analyses, plotting and curve fittings were performed using custom MATLAB scripts.

\section{Results and Discussion}

\begin{figure}[ht!]
    \centering
    \includegraphics[width=0.8\columnwidth]{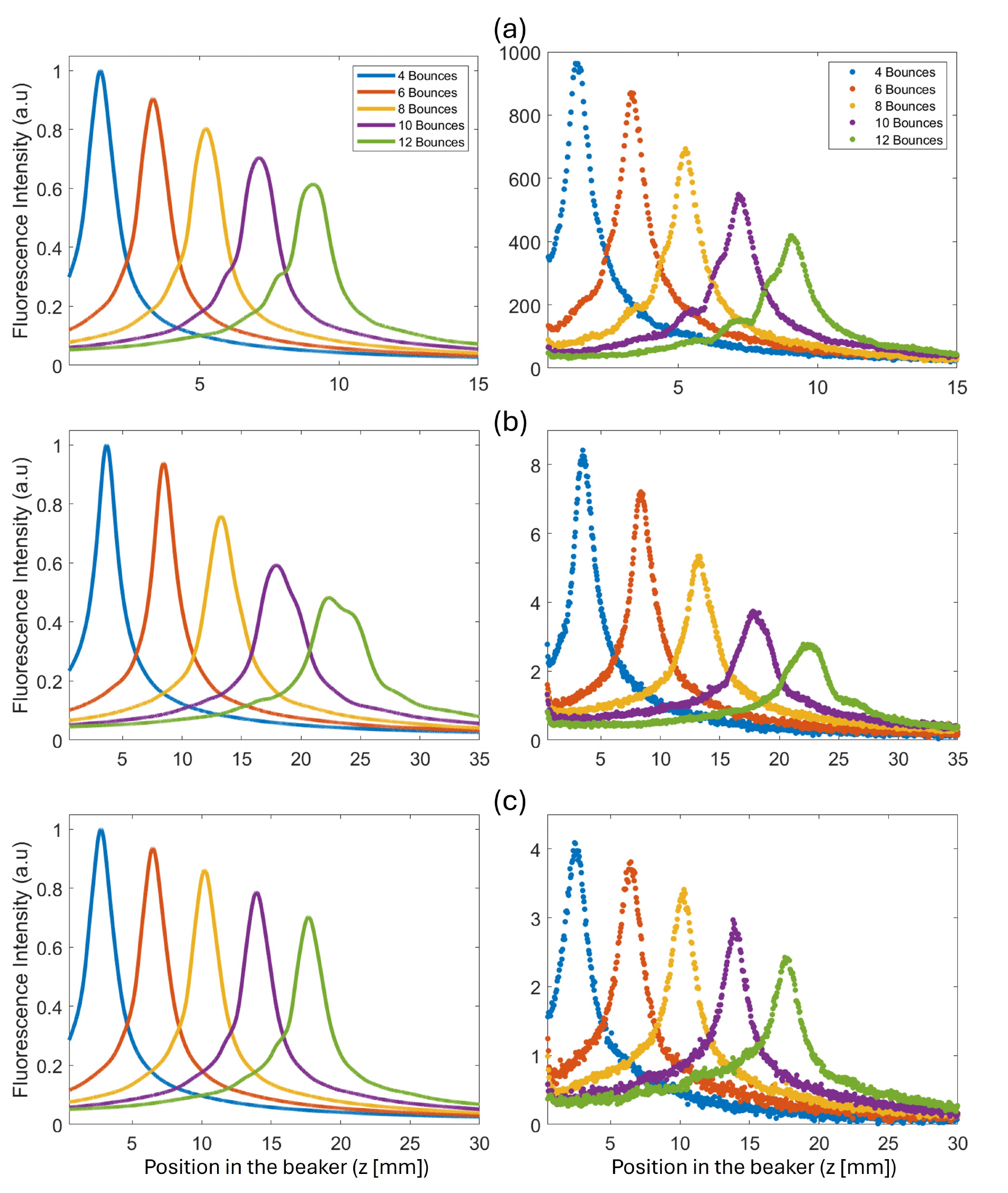}

    \caption{Measured \textbf{(right column)} and simulated \textbf{(left column)} TPA fluorescence intensity $S(z)$, plotted against the position in the liquid relative to the point of entry into the liquid. Data is shown for \textbf{(a)} chloroform, \textbf{(b)} water, and \textbf{(c)} methanol, with different colors indicating varying numbers of bounces off the CMs, as detailed in the legends.}
    \label{simulated_and_exp_s}
\end{figure}

The TPA fluorescence signal was collected for various initial negatively chirped pulse profiles with the number of bounces off the CM pair ranging from 4 to 12. A typical image illustrating the TPA fluorescence signal's dependency on the number of bounces off the CMs is shown in Fig.~\ref{experimental_steup} (b), while Fig.~\ref{simulated_and_exp_s} (right column) graphically represents the same relationship across different liquid media.

\begin{figure}[ht!]
    \centering
    \includegraphics[width=0.8\columnwidth]{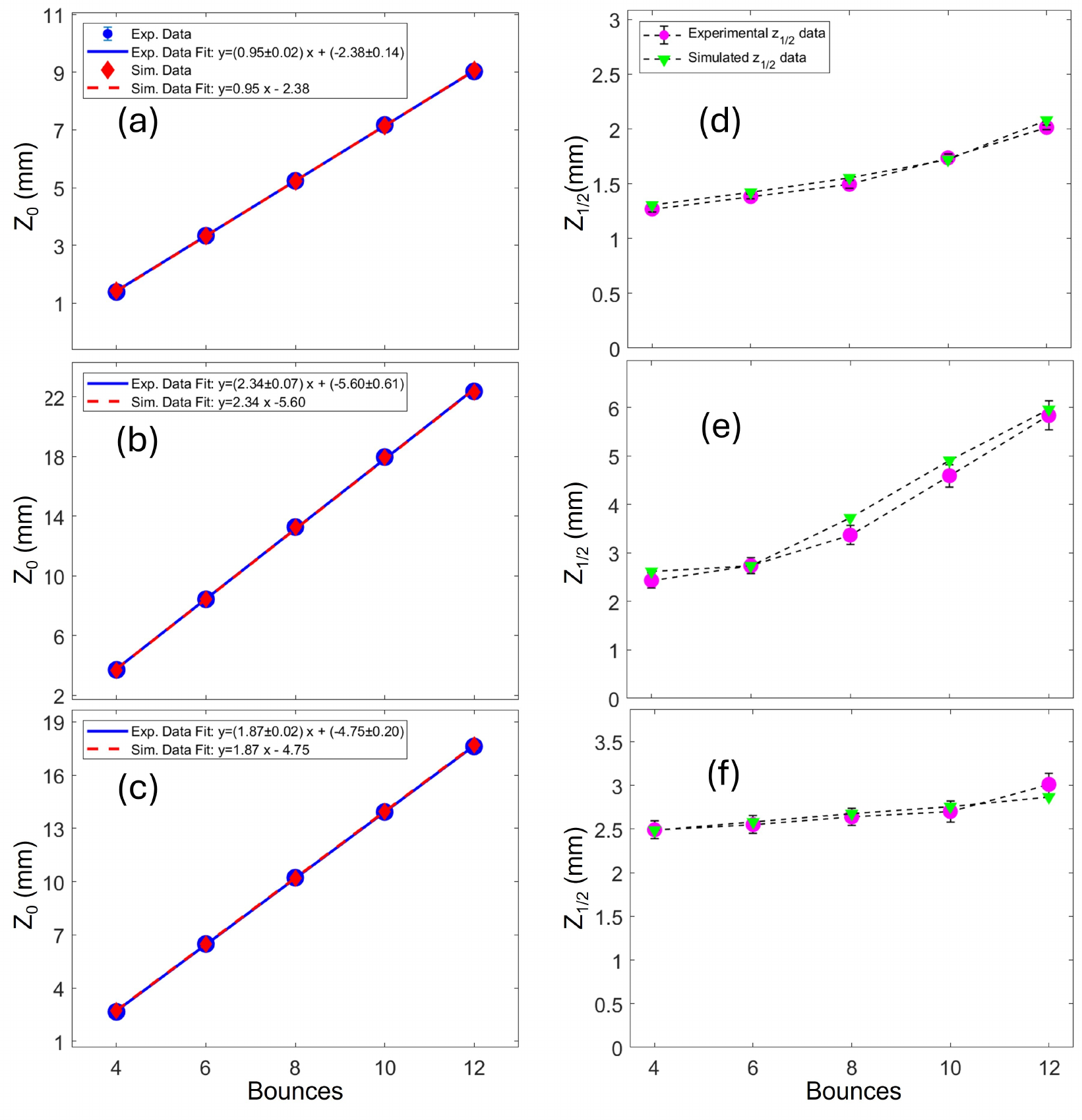}
    \caption{(a, d): chloroform, (b, e): water, and (c, f): methanol. \textbf{Left column}: Peak fluorescence intensity position $z_0$ in the beaker for different numbers of bounces off thechirped mirrors. Blue circles (dashed line) represent experimental data (fit), while red diamonds (line) represent simulated data (fit). The negative $y$-intercept values of the experimental data result from an $\text{GDD}_{\text{background}}$. \textbf{Right column}: The dependence of the number of bounces on the FWHM $z_{1/2}$ represented by experimental (pink circle) and simulated (green triangles) data. Both $z_0$ and $z_{1/2}$ are recorded in mm.}
    \label{combined_z_vs_b_and_zhalf_vs_b}
\end{figure}

\subsection{Determination of GVD}
\label{DOGVD}
To extract accurate pulse duration information from the measured TPA fluorescence signal \( S(z) \), we first determine the GVD for each liquid medium. The experimental process begins by identifying \( z_0 \), which represents the position corresponding to the peak fluorescence intensity within the liquid medium (see Fig.~\ref{simulated_and_exp_s} (right column)), as a function of the number of bounces off the CMs. This data, shown in Fig.~\ref{combined_z_vs_b_and_zhalf_vs_b} (a, b, c) using blue solid circles, facilitates the calculation of the slope \( C_\text{exp} = \frac{\Delta z_0}{\Delta \text{Bounces}} \). This slope is then incorporated into our algorithm (see Sec.~\ref{simulation} and Fig.~\ref{algorithm}) to converge the theoretical slope \( C_\text{th} \) to match the experimental value \( C_\text{exp} \), as displayed by the red diamonds in Fig.~\ref{combined_z_vs_b_and_zhalf_vs_b} (a, b, c). The values of the experimentally obtained fit parameters are given in Table \ref{tab:params}. This process yields the optimized GVD values for each specific liquid medium. The obtained GVD values for chloroform, water, and methanol are shown in Table \ref{tab:GVD_TOD}, and are found to closely align with previously reported values in the literature.

\begin{table}[ht!]
\centering
\caption{\bf Fit Parameters for $z_0$ vs Bounces for three liquid media (Fig.~\ref{combined_z_vs_b_and_zhalf_vs_b} (left column))}
\begin{tabular}{cccc}
\hline
Liquid Media & Slope (mm/Bounce) & $y$-intercept (mm)&GDD$_{\text{inherent}}$ (fs$^2$)\\
\hline
Chloroform & 0.95$\pm$0.02 & $-$2.38$\pm$0.14& 123$\pm$6\\
Water & 2.34$\pm$0.07 & $-$5.60$\pm$0.61& 123$\pm$8\\
Methanol & 1.87$\pm$0.02 & $-$4.75$\pm$0.20& 126$\pm$8\\
\hline
\end{tabular}
  \label{tab:params}
\end{table}

\begin{table}[ht!]
\centering
\caption{\bf GVD (fs$^2/$mm) and TOD (fs$^3/$mm) determined for three liquid media.}
\begin{tabular}{ccccc}
\hline
Liquid Media& GVD$_{\text{our value}}$&GVD$_{\text{prev. lit.}}$&TOD$_{\text{our value}}$&TOD$_{\text{prev. lit.}}$\\
\hline
Chloroform& 63.2$\pm$0.8 &63.5\cite{kedenburg2012linear}, 71.2\cite{chang2024refractive}& 29$\pm$2&-\\
Water& 26.0$\pm$0.4 &28.9\cite{daimon2007measurement}, 27\cite{devi2011measurement}& 39$\pm$3&34.8\cite{devi2011measurement}, 34.2\cite{coello2007group}\\
Methanol& 32.1$\pm$0.5 &33.4\cite{moutzouris2014refractive}, 32.4\cite{devi2011measurement}& 19$\pm$2&24.8\cite{devi2011measurement}\\
\hline
\end{tabular}
  \label{tab:GVD_TOD}

\end{table} 
\subsection{Determination of Background GDD}
\label{DOGDD_background} 
To align the $y$-intercepts of the curves for all three media, the background GDD was adjusted in the simulation until the intercepts matched. Taking into account the GDD contributions from air and mirrors M1 and M2 (see Fig.~\ref{experimental_steup}), the inherent GDD of the laser system may be expressed as: 
\begin{equation} 
\text{GDD}_{\text{inherent}} = \text{GDD}_{\text{background}} - \text{GDD}_{\text{air}} - \text{GDD}_{\text{M1\&M2}} 
\label{GDDinherent} 
\end{equation} 
Given that the GVD of air is 23.84 fs$^2$/m, and the laser beam traveled 0.98 m through air, the $\text{GDD}_{\text{air}}$ is calculated as 23.36 fs$^2$. The GDD from mirrors M1 and M2 ($\text{GDD}_{\text{M1\&M2}} $) was specified by the manufacturer as $-$2 fs$^2$. The resulting values of $\text{GDD}_{\text{inherent}}$, presented in Table \ref{tab:params}, remain consistent ($124 \pm 4$ fs$^2$) across all three media, suggesting that this term arises from the inherent GDD of the laser system itself, rather than from any medium-specific contributions.

\subsection{Determination of TOD}
\label{DOTOD}
The increasing widths of the measured TPA fluorescence signal \( S(z) \) curves with the number of bounces, as shown in Fig.~\ref{simulated_and_exp_s} (right column), suggest contributions from TOD. This trend is further highlighted by the solid pink circles in Fig.~\ref{combined_z_vs_b_and_zhalf_vs_b} (right column), where the FWHM (or \( z_{1/2} \)) values increases with the number of bounces. The potential sources of TOD in our experiment include the liquid medium, the CMs (see Fig.~\ref{spectra_GD_combined}), and the laser source itself. Since we utilized actual manufacturer-provided data from Thorlabs for the CMs in our simulations, and assuming the inherent TOD of the laser is negligible, the liquid medium remains the most likely source of TOD.

To account for this, the algorithm iteratively adjusted the TOD values until the simulated \( S(z) \) curves matched the experimental \( S(z) \) curve shapes for each medium, as shown in Fig.~\ref{simulated_and_exp_s} (left column). As a result, the simulated \( z_{1/2} \) vs. bounces data, represented by green triangles in Fig.~\ref{combined_z_vs_b_and_zhalf_vs_b} (right column), closely aligned with the experimentally obtained values (pink circles). The TOD values obtained from this process are listed in Table~\ref{tab:GVD_TOD} and are consistent with values reported in previous literature.

\subsection{Determination of Pulse Duration}

\begin{figure}[ht!]
    \centering
    \includegraphics[width=0.8\columnwidth]{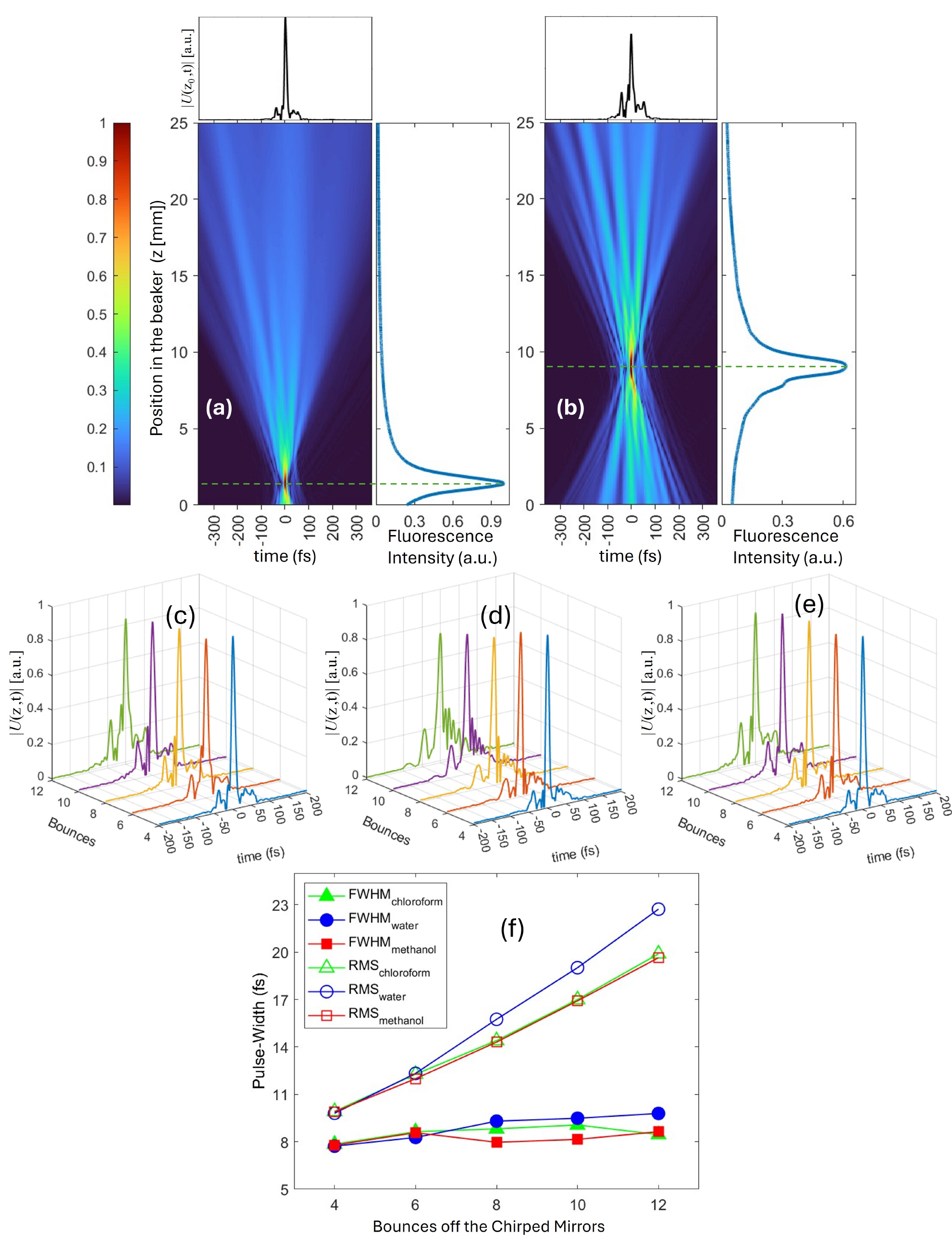}
\caption{The relationship between TPA fluorescence signal (blue trace) and the absolute pulse amplitude profile is illustrated for chloroform, presented for (a) 4 bounces and (b) 12 bounces off the CMs. The green dashed line indicates the position $z_0$ where the pulse is shortest in the liquid, with the corresponding temporal pulse amplitude at this location depicted by the black curve at the top. Panels (c), (d), and (e) show the simulated absolute pulse amplitude profiles $|U(z,t)|$ for chloroform, water, and methanol, respectively. Panel (f) presents a comparison of two pulse duration measurements, root-mean-square (RMS) and full-width-at-half-maximum (FWHM), derived from $|U(z,t)|^2$ for each medium, plotted as a function of the number of bounces after the pulse has traveled a distance $z_0$ in the media.}
    \label{pulse_fluorescence}
\end{figure}

To illustrate how we extracted pulse information from the TPA fluorescence signal, consider a pulse entering the beaker. As it propagates through the liquid, it gives rise to a fluorescence intensity profile as a function of propagation distance $z$ with a maximum located at $z_0$ (green dashed line in Figs.~\ref{pulse_fluorescence} (a) and (b)). The width of the fluorescence profile is not directly indicative of the pulse duration itself. A $\sim$10 fs pulse traveling at \( c/n \approx 2 \times 10^8 \, \text{m/s} \) has a spatial extension of $\sim $2 \(\mu\)m, a negligible size compared to the $\sim$50 mm beaker length. To visualize the pulse duration changes as it propagates in the liquid, the calculated temporal pulse shape where time is indicated perpendicular to $z$, is shown as a function of $z$ in Figs.~\ref{pulse_fluorescence} (a) and (b). The integration over time (Eq.~(\ref{4})), gives the fluorescent intensity profile. In panel (a), we use only four bounces off the CMs, which causes the pulse to reach a minimum duration soon after it has entered the beaker (horizontal green dashed line). In panel (b), we let the pulse bounce off the CMs 12 times. This has two effects on the fluorescence signal (blue trace). First, the added negative chirp now requires more distance in the liquid to be canceled, causing the brightest spot to appear deeper in the beaker (horizontal green dashed line). Second, the pulse gets contaminated with higher-order phases. This is due to the larger number of bounces, with each bounce imprinting a higher-order phase onto the beam. In addition, the pulse acquires higher-order phases from the liquid due to its longer propagation in the beaker before it reaches a minimum duration. As a result, the temporal pulse shape becomes more distorted and the width of the fluorescence signal increases. Note that the relatively small increase in pulse duration ($\sim$20\%) is accompanied with a larger broadening of fluorescence intensity profile ($\sim$100\%), as the pulse stays short over a longer length in the beaker.

More specifically, for each liquid medium, we plotted the temporal pulse shape of the amplitude $|U(z,t)|$ for different numbers of bounces (Figs.~\ref{pulse_fluorescence} (c,d,e)) as occurring at the locations in the medium where the fluorescence is brightest ($z=z_0$, compare green dashed lines in Figs.\ref{pulse_fluorescence} (a) and (b)). In all cases, the central peak remains sharp, but a pedestal due to increasing amounts of higher-order phases becomes more noticeable when more CMs are used, forcing the the pulse to travel deeper into the liquid to achieve focusing. Hence, the FWHM is relatively insensitive to the use of additional CMs, whereas the RMS duration increases as can be seen in Fig.~\ref{pulse_fluorescence} (f) for all three liquids.

\begin{table}[ht!]
\centering
\caption{\bf Mean Pulse Durations (FWHM and RMS) as a Function of Bounces in Various Liquid Media.}
\begin{tabular}{ccc}
\hline

Bounces&Mean FWHM (fs)&Mean RMS (fs)\\
\hline
4&7.78$\pm$0.05&9.87$\pm$0.05\\
6&8.49$\pm$0.16&12.20$\pm$0.15\\
8&8.70$\pm$0.55&14.82$\pm$0.66\\
10&8.90$\pm$0.56&17.64$\pm$0.97\\
12&8.96$\pm$0.59&20.75$\pm$1.39\\
Transform-Limited$^\textit{*}$&6.75$^\textit{*}$&7.81$^\textit{*}$\\
\hline
\end{tabular}
  \label{tab:FWHM&RMS}
$^\textit{*} \text{computed from } |U(z_0,t)|^2 = |\mathcal{F}^{-1}\{\tilde{U}(0,\omega)\}|^2.$
\end{table}

\begin{figure}[ht!]
    \centering
    \includegraphics[width=0.4\columnwidth]{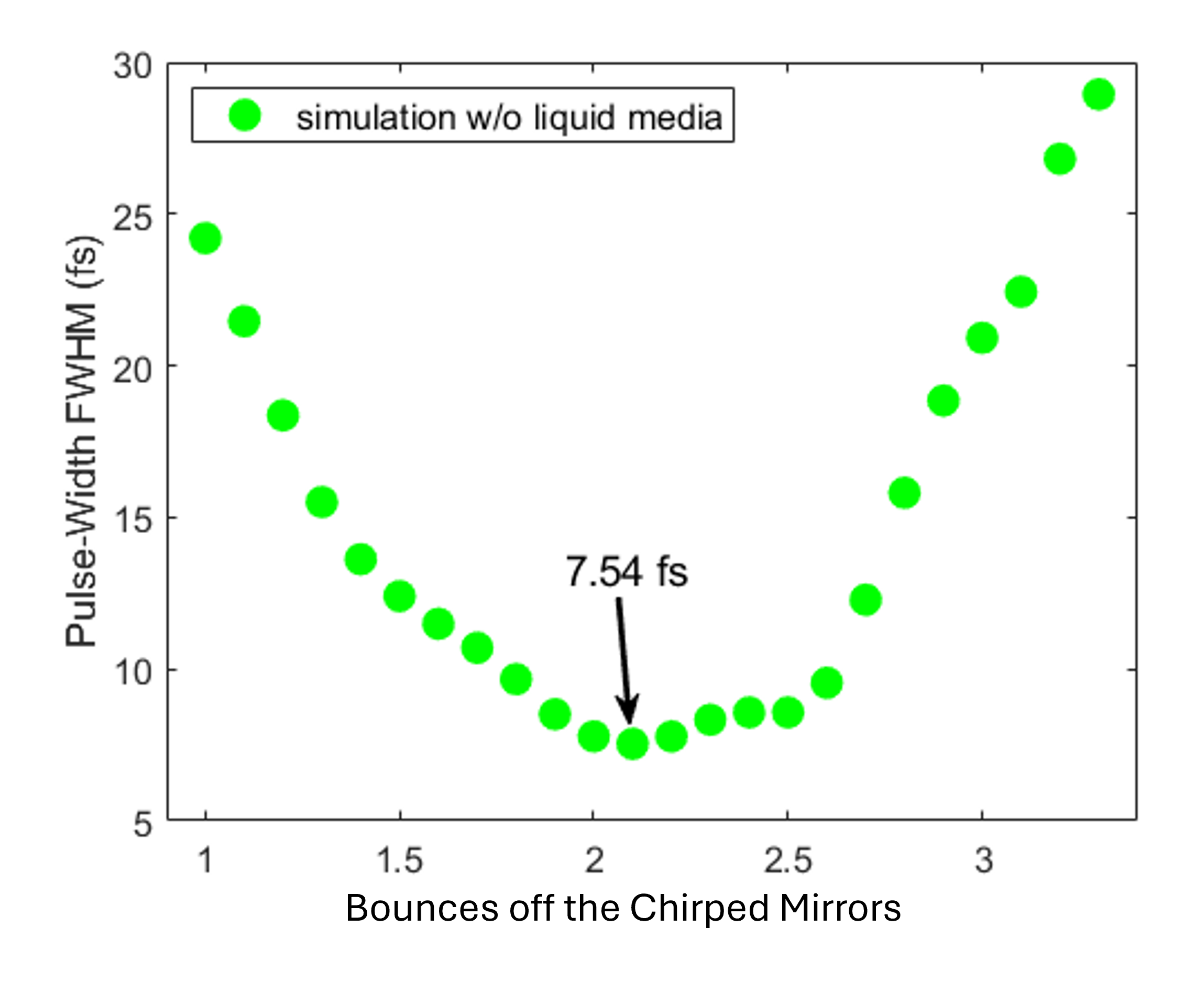}
\caption{The shortest FWHM pulse duration of our laser, observed at a projected value of 2.1 bounces, indicating optimal compensation of the laser's inherent GDD by our CMs.}
    \label{illustration}
\end{figure}

Table~\ref{tab:FWHM&RMS} summarizes the pulse duration, presenting averaged FWHM and RMS values for all three liquid media across various bounces off the CMs. Even at just 4 bounces, the FWHM pulse duration is $7.78\pm0.05$ fs, which is higher than the transform-limited value of 6.75 fs. Notably, the ratio of RMS to FWHM is at least four times greater than that of a Gaussian pulse profile, for which FWHM $\approx$ 2.35$\times$RMS. This represents the deviation of our laser's pulse profile from the standard Gaussian one.

Fig.~\ref{illustration} depicts a simulated scenario where the laser's inherent GDD of \( 124 \pm 4 \, \text{fs}^2 \) is compensated using our CMs, with no additional media in the laser path. A minimum pulse duration of 7.54 fs is achieved at an projected value of 2.1 bounces off the CMs.

\begin{figure}[ht!]
    \centering
    \includegraphics[width=0.8\columnwidth]{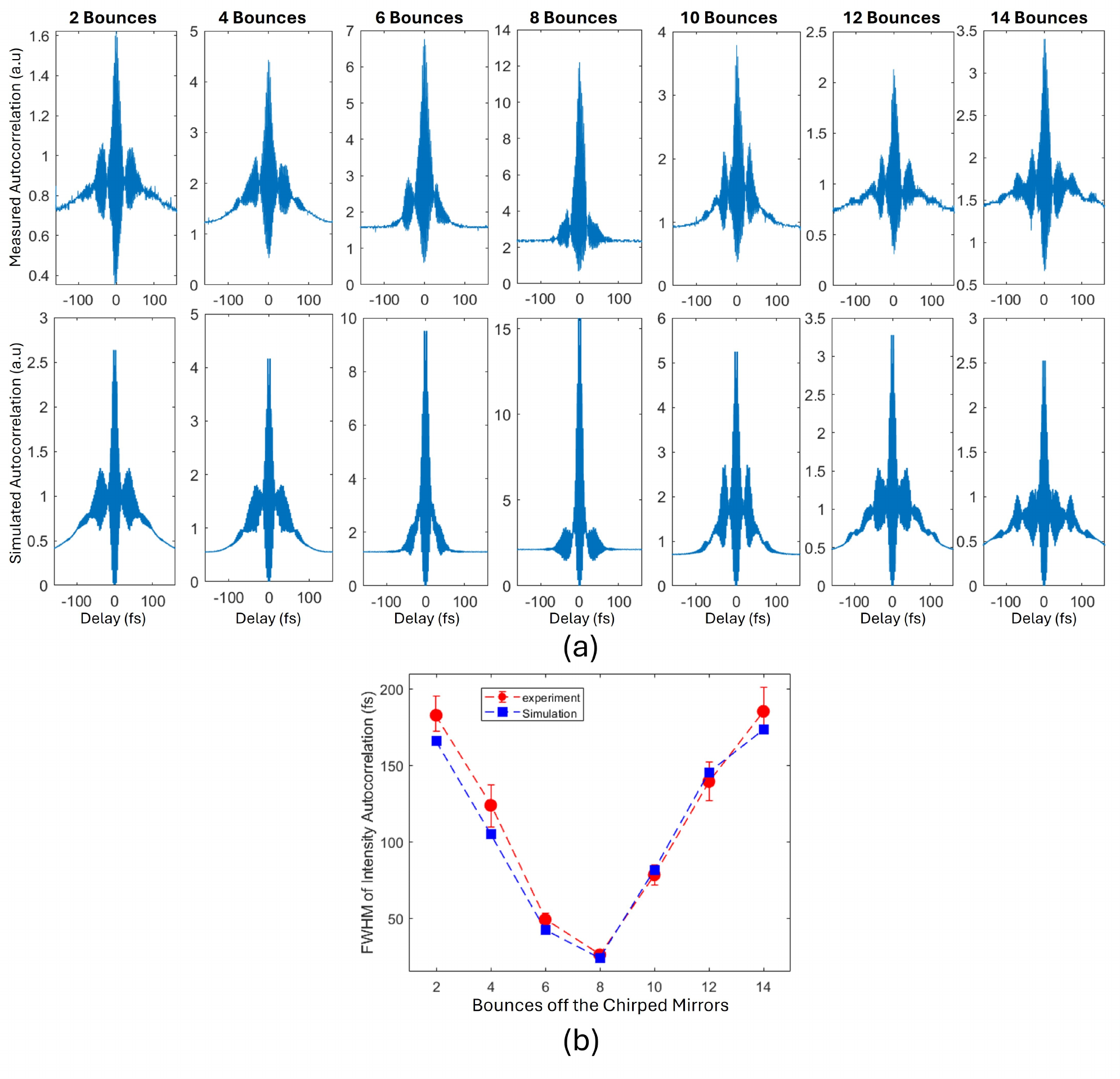}
\caption{(a) \textbf{Top row:} Measured autocorrelation traces using Thorlabs' FSAC Interferometric Autocorrelator for varying numbers of bounces (2 to 14) off the CMs. \textbf{Bottom row:} Simulated autocorrelation traces under identical conditions, accounting for the inherent GDD of the laser, the air medium, and the manufacturer-provided GDD and TOD of the FSAC Autocorrelator. (b) FWHM of the intensity autocorrelation plotted as a function of bounces off the CMs.}
    \label{combined_autocorrelator}
\end{figure}

\subsection{Independent Verification}

To further validate our simulation results, we performed interferometric autocorrelation measurements of our laser with a non-linear detector. Fig.~\ref{combined_autocorrelator} (a) shows the measured (top row) and simulated (bottom row) autocorrelation signals for different numbers of bounces. Agreement is observed in the structures of the autocorrelation traces obtained experimentally and through our simulation. Fig.~\ref{combined_autocorrelator} (b) demonstrates the close match between the FWHM of the intensity autocorrelation for both measured and simulated data. The best match between the simulated and experimental autocorrelations was achieved when the value of $\text{GDD}_{\text{inherent}}$ was adjusted to 121 fs$^2$ in our simulation, which falls within the error bars of the previously determined $\text{GDD}_{\text{inherent}}$ from our earlier experiment (see Sec.~\ref{GDDinherent}). Therefore, the autocorrelation measurements independently support our determined value of $\text{GDD}_{\text{inherent}}$ and the retrieved pulse information, thereby reinforcing the validity of our simulation code.

\section{Conclusion}
In conclusion, this research successfully demonstrates a novel methodology for characterizing the pulse duration of ultrafast laser using TPA fluorescence and computational simulations. Our key findings include:
\begin{itemize}
    \item A retrieved pulse duration of 7.54 fs FWHM for the Octavius Femtosecond laser in media without significant TOD, closely matching the expected transform-limited pulse of 6.7 fs.
    \item Numerical simulations effectively captured complex dispersion effects, including GVD and TOD, while revealing the inherent GDD of the laser.
    \item The approach provided valuable insights into the laser’s non-Gaussian output and phase distortions introduced by CMs.
    \item Our approach also enables us to determine the TOD of the liquids water, methanol, and chloroform.  
\end{itemize}

This combined experimental and computational strategy presents a practical alternative to conventional techniques, significantly reducing the need for precise alignment while maintaining high accuracy. Additionally, this method can accurately detect and quantify the GDD introduced by optical elements such as half-wave plates, polarizers, and optical windows, which is crucial for precision optics. Future work could extend this methodology to explore additional laser systems and media, enhancing our understanding of ultrafast optical phenomena and their applications in diverse fields.

\section*{Acknowledgments}
This material is in part based upon work supported by the National Science Foundation under Grant No. NSF PHY-2207697 and the University of Nebraska Program of Excellence (PoE). Special thanks to Mr. Arjun Mohanan for his flexibility and cooperation in sharing the laser beam time, enabling us to complete the experimental measurements.
\section*{Disclosures}
The authors declare no conflicts of interest. ASML was not involved in this research.
\section*{Data availability} Data underlying the results presented in this paper are not publicly available at this time but may be obtained from the authors upon reasonable request.
\bibliography{main}
\end{document}